\documentclass[aps,preprintnumbers]{revtex4}
\usepackage{epsfig}

\newcommand{\beq}{\begin{equation}}
\newcommand{\beqn}{\begin{eqnarray}}
\newcommand{\eeq}{\end{equation}}
\newcommand{\eeqn}{\end{eqnarray}}
\begin{document}

\preprint{MIT-CTP-3594}

\title{Inflationary Cosmology: \\
Exploring the Universe from the Smallest to the Largest Scales}

\author{Alan H. Guth}
\email[Email: ]{guth@ctp.mit.edu}
\author{David I. Kaiser}
\email[Email: ]{dikaiser@mit.edu}
\affiliation{Center for Theoretical
Physics, Laboratory for Nuclear Science and Department of
Physics,\\ Massachusetts Institute of Technology, Cambridge,
Massachusetts 02139}
\date{16 February 2005}

\begin{abstract}
Understanding the behavior of the universe at large depends
critically on insights about the smallest units of matter and
their fundamental interactions.  Inflationary cosmology is a
highly successful framework for exploring these interconnections
between particle physics and gravitation.  Inflation makes
several predictions about the present state of the universe ---
such as its overall shape, large-scale smoothness, and
smaller-scale structure --- which are being tested to
unprecedented accuracy by a new generation of astronomical
measurements.  The agreement between these predictions and the
latest observations is extremely promising.  Meanwhile,
physicists are busy trying to understand inflation's ultimate
implications for the nature of matter, energy, and spacetime.

\end{abstract}

\maketitle

\baselineskip 16pt

The scientific community is celebrating the International Year of
Physics in 2005, honoring the centennial of Albert Einstein's
most important year of scientific innovation.  In the span of
just a few months during 1905 Einstein introduced key notions
that would dramatically change our understanding of matter and
energy as well as the nature of space and time.  The centennial
of these seminal developments offers an enticing opportunity to
take stock of how scientists think about these issues today.  We
focus in particular on recent developments in the field of
inflationary cosmology, which draws on a blend of concepts from
particle physics and gravitation.  The last few years have been a
remarkably exciting time for cosmology, with new observations of
unprecedented accuracy yielding many surprises.  Einstein's
legacy is flourishing in the early 21st century.

Inflation was invented a quarter of a century ago, and has become
a central ingredient of current cosmological research. 
Describing dramatic events in the earliest history of our
universe, inflationary models generically predict that our
universe today should have several distinct features --- features
that are currently being tested by the new generation of
high-precision astronomical measurements.  Even as inflation
passes more and more stringent empirical tests, theorists
continue to explore broader features and implications, such as
what might have come before an inflationary epoch, how inflation
might have ended within our observable universe, and how
inflation might arise in the context of our latest understanding
of the structure of space, time, and matter.

Particle theory has been changing rapidly, and these theoretical
developments have provided just as important a spur to
inflationary cosmology as have the new observations.  During the
1960s and 70s, particle physicists discovered that if they
neglected gravity, they could construct highly successful
descriptions of three out of the four basic forces in the
universe: electromagnetism and the strong and weak nuclear
forces.  The ``standard model of particle physics,'' describing
these three forces, was formulated within the framework of
quantum field theory, the physicist's quantum-mechanical
description of subatomic matter.  Inflationary cosmology was
likewise first formulated in terms of quantum field theory.  Now,
however, despite (or perhaps because of) the spectacular
experimental success of the standard model, the major thrust of
particle physics research is aimed at moving beyond it.

For all its successes, the standard model says nothing at all
about the fourth force: gravity.  For more than 50 years
physicists have sought ways to incorporate gravity within a
quantum-mechanical framework, initially with no success.  But for
the past 25 or more years, an ever-growing group of theoretical
physicists has been pursuing superstring theory as the bright
hope for solving this problem.  To accomplish this task, however,
string theorists have been forced to introduce many novel
departures from conventional ideas about fundamental forces and
the nature of the universe.  For one thing, string theory
stipulates that the basic units of matter are not pointlike
particles (as treated by quantum field theory), but rather
one-dimensional extended objects, or strings.  Moreover, in order
to be mathematically self-consistent, string theories require the
existence of several additional spatial dimensions.  Whereas our
observable universe seems to contain one timelike dimension and
three spatial dimensions --- height, width, and depth --- string
theory postulates that our universe actually contains at least
six additional spatial dimensions, each at right angles to the
others and yet somehow hidden from view. 

For measurements at low energies, string theory should behave
effectively like a quantum field theory, reproducing the
successes of the standard model of particle physics.  Yet the
interface between cosmology and string theory has been a lively
frontier.  For example, some theorists have been constructing
inflationary models for our universe that make use of the extra
dimensions that string theory introduces.  Others have been
studying the string theory underpinnings for inflationary models,
exploring such topics as the nature of vacuum states and the
question of their uniqueness. As we will see, inflation continues
to occupy a central place in cosmological research, even as its
relation to fundamental particle physics continues to evolve.

\section{Inflationary Basics}

According to inflationary cosmology
\cite{Guth1981,Linde1982,AlbrechtSteinhardt1982},
the universe expanded exponentially quickly for a fraction of a
second very early in its history --- growing from a patch as
small as $10^{-26}$ m, one hundred billion times smaller than a
proton, to macroscopic scales on the order of a meter, all within
about $10^{-35}$ s --- before slowing down to the more
stately rate of expansion that has characterized the universe's
behavior ever since.  The driving force behind this dramatic
growth, strangely enough, was gravity.  [For technical
introductions to inflationary cosmology, see
\cite{Linde1990,KolbTurner1990,LiddleLyth2000}; a more popular
description may be found in \cite{GuthBook}.]  Although this
might sound like hopeless (or, depending on one's inclinations,
interesting) speculation, in fact inflationary cosmology leads to
several quantitative predictions about the present behavior of
our universe --- predictions that are being tested to
unprecedented accuracy by a new generation of observational
techniques.  So far the agreement has been excellent.

How could gravity drive the universal repulsion during inflation?
The key to this rapid expansion is that in Einstein's general
relativity (physicists' reigning description of gravity), the
gravitational field couples both to mass-energy (where mass and
energy are interchangeable thanks to Einstein's $E = mc^2$) and
to pressure, rather than to mass alone.  In the simplest
scenario, in which at least the observable portion of our
universe can be approximated as being homogeneous and isotropic
--- that is, having no preferred locations or directions ---
Einstein's gravitational equations give a particularly simple
result.  The expansion of the universe may be described by
introducing a time-dependent ``scale factor,'' $a(t)$, with the
separation between any two objects in the universe being
proportional to $a(t)$.  Einstein's equations prescribe how this
scale factor will evolve over time, $t$.  The rate of
acceleration is proportional to the density of mass-energy in the
universe, $\rho$, plus three times its pressure, $p$: $d^2 a /
dt^2 = - 4\pi G (\rho + 3p) a / 3$, where $G$ is Newton's
gravitational constant (and we use units for which the speed of
light $c=1$). The minus sign is important: ordinary matter under
ordinary circumstances has both positive mass-energy density and
positive (or zero) pressure, so that $(\rho + 3p) > 0$.  In this
case, gravity acts as we would expect it to: All of the matter in
the universe tends to attract all of the other matter, causing
the expansion of the universe as a whole to slow down.

The crucial idea behind inflation is that matter can behave
rather differently from this familiar pattern.  Ideas from
particle physics suggest that the universe is permeated by scalar
fields, such as the Higgs field of the standard model of particle
physics, or its more exotic generalizations.  (A scalar field
takes exactly one value at every point in space and time.  For
example, one could measure the temperature at every position in a
room and repeat the measurements over time, and represent the
measurements by a scalar field, $T$, of temperature.  Electric
and magnetic fields are vector fields, which carry three distinct
components at every point in space and time: the field in the $x$
direction, in the $y$ direction, and in the $z$ direction. 
Scalar fields are introduced in particle physics to describe
certain kinds of particles, just as photons are described in
quantum field theories in terms of electromagnetic fields.) These
scalar fields can exist in a special state, having a high energy
density that cannot be rapidly lowered, such as the arrangement
labeled (a) in Fig.~1.  Such a state is called a ``false
vacuum.'' Particle physicists use the word ``vacuum'' to denote
the state of lowest energy.  ``False vacua'' are only metastable,
not the true states of lowest possible energy.

In the early universe, a scalar field in such a false vacuum
state can dominate all the contributions to the total mass-energy
density, $\rho$.  During this period, $\rho$ remains nearly
constant, even as the volume of the universe expands rapidly:
$\rho \cong \rho_f = {\rm constant}$. This is quite different
from the density of ordinary matter, which decreases when the
volume of its container increases.  Moreover, the first law of
thermodynamics, in the context of general relativity, implies
that if $\rho \cong \rho_f$ while the universe expands, then the
equation of state for this special state of matter must be $p
\cong - \rho_f$, a negative pressure.  This yields $d^2 a / dt^2
= 8\pi G\rho_f a /3$: Rather than slowing down, the cosmic
expansion rate will grow rapidly, driven by the negative pressure
created by this special state of matter.  Under these
circumstances, the scale factor grows as $a \propto e^{Ht}$,
where the Hubble parameter, $H \equiv a^{-1} da / dt$, which
measures the universe's rate of expansion, assumes the constant
value, $H \cong \sqrt{8\pi G \rho_f / 3}$. The universe expands
exponentially until the scalar field rolls to near the bottom of
the hill in the potential energy diagram.
\begin{figure}
\centerline{\epsfig{file=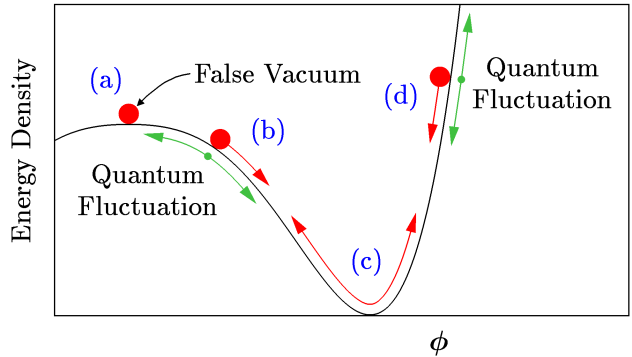}}
\vspace{10pt}
\caption{\small In simple inflationary models, the universe at
early times is dominated by the potential energy density of a
scalar field, $\phi$.  Red arrows show the classical motion of
$\phi$.  When $\phi$ is near region (a), the energy density will
remain nearly constant, \hbox{$\rho \protect\cong \rho_f$}, even
as the universe expands.  Moreover, cosmic expansion acts like a
frictional drag, slowing the motion of $\phi$: Even near regions
(b) and (d), $\phi$ behaves more like a marble moving in a bowl
of molasses, slowly creeping down the side of its potential,
rather than like a marble sliding down the inside of a polished
bowl.  During this period of ``slow roll,'' $\rho$ remains nearly
constant.  Only after $\phi$ has slid most of the way down its
potential will it begin to oscillate around its minimum, as in
region (c), ending inflation.}
\end{figure}

What supplies the energy for this gigantic expansion? The answer,
surprisingly, is that no energy is needed \cite{GuthBook}. 
Physicists have known since the 1930s \cite{Tolman1932} that the
gravitational field carries {\em negative} potential energy
density.  As vast quantities of matter are produced during
inflation, a vast amount of negative potential energy
materializes in the gravitational field that fills the
ever-enlarging region of space.  The total energy remains
constant, and very small, and possibly exactly equal to zero.

There are now dozens of models that lead to this generic
inflationary behavior, featuring an equation of state, $p \cong -
\rho$, during the early universe \cite{Linde1990,LythRiotto1999}. 
This entire family of models, moreover, leads to several main
predictions about today's universe.  First, our observable
universe should be spatially flat.  Einstein's general relativity
allows for all kinds of curved (or ``non-Euclidean'') spacetimes. 
Homogeneous and isotropic spacetimes fall into three classes
(Fig.~2), depending on the value of the mass-energy density,
$\rho$.  If $\rho > \rho_c$, where $\rho_c \equiv 3 H^2 / (8\pi
G)$, then Einstein's equations imply that the spacetime will be
positively curved, or closed (akin to the two-dimensional surface
of a sphere); parallel lines will intersect, and the interior
angles of a triangle will always add up to more than $180^\circ$. 
If $\rho < \rho_c$, the spacetime will be negatively curved, or
open (akin to the two-dimensional surface of a saddle); parallel
lines will diverge and triangles will sum to less than
$180^\circ$.  Only if $\rho = \rho_c$ will spacetime be spatially
flat (akin to an ordinary two-dimensional flat surface); in this
case, all of the usual rules of Euclidean geometry apply. 
Cosmologists use the letter $\Omega$ to designate the ratio of
the actual mass-energy density in the universe to this critical
value: $\Omega \equiv \rho / \rho_c$.  Although general
relativity allows any value for this ratio, inflation predicts
that $\Omega = 1$ within our observable universe to extremely
high accuracy.  Until recently, uncertainties in the measurement
of $\Omega$ allowed any value in the wide range, $0.1 \leq \Omega
\leq 2$, with many observations seeming to favor $\Omega \cong
0.3$. A new generation of detectors, however, has dramatically
changed the situation.  The latest observations, combining data
from the Wilkinson Microwave Anisotropy Probe (WMAP), the Sloan
Digital Sky Survey (SDSS), and observations of type Ia
supernovae, have measured $\Omega = 1.012^{+0.018}_{-0.022}$
\cite{Tegmark2004} --- an amazing match between prediction and
observation.
\begin{figure}
\centerline{\epsfig{file=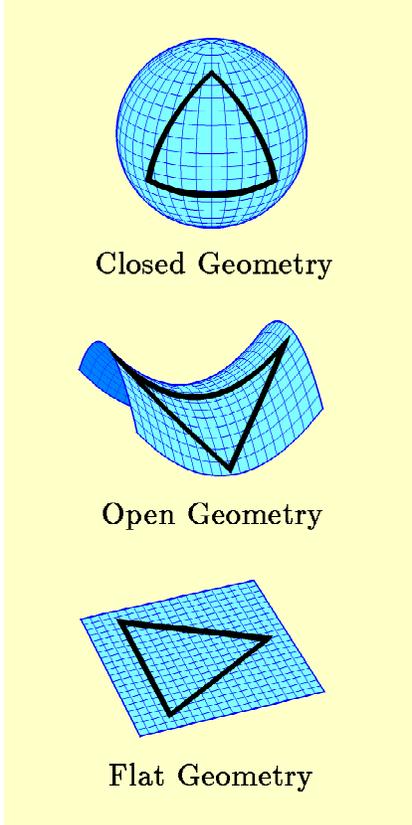}}
\vspace{10pt}
\caption{\small According to general relativity, spacetime may be
warped or curved, depending on the density of mass-energy. 
Inflation predicts that our observable universe should be
spatially flat to very high accuracy.}
\end{figure}

In fact, inflation offers a simple explanation for why the
universe should be so flat today.  In the standard big bang
cosmology (without inflation), $\Omega = 1$ is an {\em unstable}
solution: if $\Omega$ were ever-so-slightly less than 1 at an
early time, then it would rapidly slide toward 0.  For example,
if $\Omega$ were $0.9$ at 1 s after the big bang, it would
be only $10^{-14}$ today.  If $\Omega$ were 1.1 at $t = 1$ s,
then it would have grown so quickly that the universe would have
recollapsed just 45 s later.  In inflationary models, on
the other hand, any original curvature of the early universe
would have been stretched out to near-flatness as the universe
underwent its rapid expansion (Fig.~3).  Quantitatively, $|\Omega
- 1| \propto 1/(aH)^2$, so that while $H \cong {\rm constant}$
and $a \propto e^{Ht}$ during the inflationary epoch, $\Omega$
gets driven rapidly to 1.
\begin{figure}
\centerline{\epsfig{file=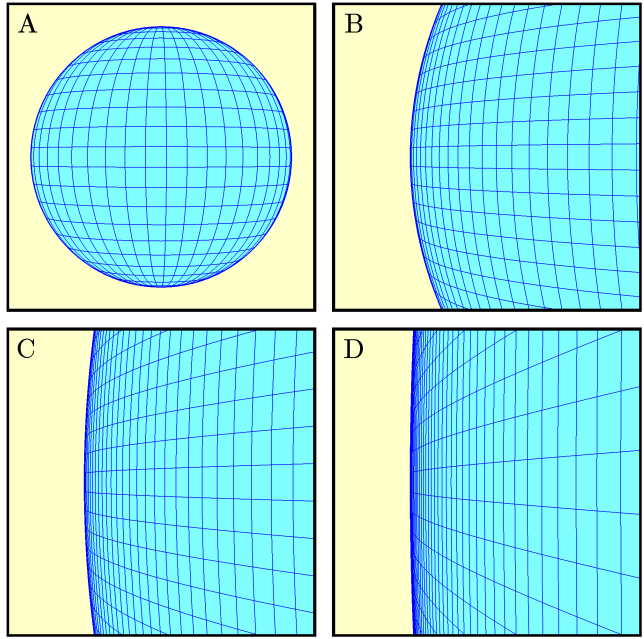}}
\vspace{10pt}
\caption{\small (A to D) The expanding sphere illustrates the
solution to the flatness problem in inflationary cosmology.  As
the sphere becomes larger, its surface becomes more and more
flat.  In the same way, the exponential expansion of spacetime
during inflation causes it to become spatially flat.}
\end{figure}

The second main prediction of inflation is that the presently
observed universe should be remarkably smooth and homogeneous on
the largest astronomical scales.  This, too, has been measured to
extraordinary accuracy during the past decades.  Starting in the
1960s, Earth-bound, balloon-borne, and now satellite detectors
have measured the cosmic microwave background (CMB) radiation, a
thermal bath of photons that fills the sky.  The photons today
have a frequency that corresponds to a temperature of 2.728 K
\cite{Fixsen1996}.  These photons were released 
$\sim$400,000 years after the inflationary epoch, when the
universe had cooled to a low enough temperature that would allow
stable (and electrically neutral) atoms to form.  Before that
time, the ambient temperature of matter in the universe was so
high that would-be atoms were broken up by high-energy photons as
soon as they formed, so that the photons were effectively
trapped, constantly colliding into electrically charged matter. 
Since stable atoms formed, however, the CMB photons have been
streaming freely.  Their temperature today is terrifically
uniform: After adjusting for the Earth's motion, CMB photons
measured from any direction in the sky have the same temperature
to one part in $10^5$ \cite{WMAP2003}.

Without inflation, this large-scale smoothness appears quite
puzzling.  According to ordinary (noninflationary) big bang
cosmology, these photons should never have had a chance to come
to thermal equilibrium: The regions in the sky from which they
were released would have been about 100 times farther apart than
even light could have traveled between the time of the big bang
and the time of the photons' release
\cite{Guth1981,Linde1990,KolbTurner1990,LiddleLyth2000}.  Much
like the flatness problem, inflation provides a simple and
generic reason for the observed homogeneity of the CMB: Today's
observable universe originated from a {\em much} smaller region
than that in the noninflationary scenarios.  This much-smaller
patch could easily have become smooth before inflation began. 
Inflation would then stretch this small homogeneous region to
encompass the entire observable universe.

A third major prediction of inflationary cosmology is that there
should be tiny departures from this strict large-scale
smoothness, and that these ripples (or ``perturbations'') should
have a characteristic spectrum.  Today these ripples can be seen
directly as fluctuations in the CMB\null.  Although the ripples
are believed to be responsible for the grandest structures of the
universe --- galaxies, superclusters, and giant voids --- in
inflationary models they arise from quantum fluctuations, usually
important only on atomic scales or smaller.  The field $\phi$
that drives inflation, like all quantum fields, undergoes quantum
fluctuations in accord with the Heisenberg uncertainty principle.
During inflation these quantum fluctuations are stretched
proportionally to $a(t)$, rapidly growing to macroscopic scales. 
The result: a set of nearly scale-invariant perturbations
extending over a huge range of wavelengths \cite{DensPert}. 
Cosmologists parameterize the spectrum of primordial
perturbations by a spectral index, $n_s$.  A scale-invariant
spectrum would have $n_s = 1.00$; inflationary models generically
predict $n_s = 1$ to within $\sim 10\%$. The latest measurements
of these perturbations by WMAP and SDSS reveal $n_s =
0.977^{+0.039}_{-0.025}$ \cite{Tegmark2004}.
\begin{figure}
\centerline{\epsfig{file=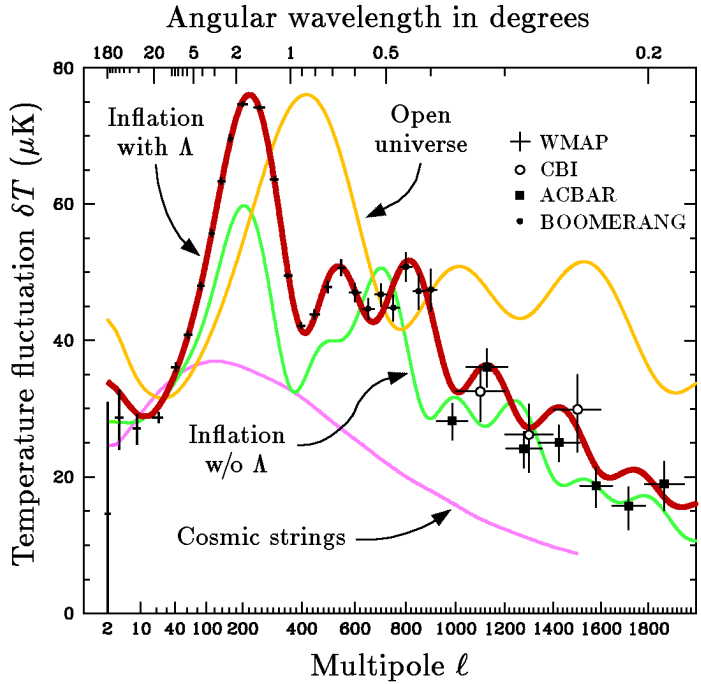}}
\vspace{10pt}
\caption{\small Comparison of the latest observational
measurements of the temperature fluctuations in the CMB with
several theoretical models, as described in the text.  The
temperature pattern on the sky is expanded in multipoles (i.e.,
spherical harmonics), and the intensity is plotted as a function
of the multipole number $\ell$.  Roughly speaking, each multipole
$\ell$ corresponds to ripples with an angular wavelength of
$360^\circ/\ell$.}
\end{figure}

Until recently, astronomers were aware of several cosmological
models that were consistent with the known data: an open
universe, with $\Omega \cong 0.3$; an inflationary universe with
considerable dark energy ($\Lambda$); an inflationary universe
without $\Lambda$; and a universe in which the primordial
perturbations arose from topological defects such as cosmic
strings.  Dark energy \cite{Kirshner} is a form of matter with
negative pressure that is currently believed to contribute more
than $70\%$ of the total energy of the observed universe. Cosmic
strings are long, narrow filaments hypothesized to be scattered
throughout space, remnants of a symmetry-breaking phase
transition in the early universe \cite{Vilenkin,Pen97}.  [Cosmic
strings are topologically nontrivial configurations of fields
which should not be confused with the fundamental strings of
superstring theory.  The latter are usually believed to have
lengths on the order of $10^{-35}$ m, although for some
compactifications these strings might also have astronomical
lengths \cite{PolchinskiFD}.] Each of these models leads to a
distinctive pattern of resonant oscillations in the early
universe, which can be probed today through its imprint on the
CMB\null.  As can be seen in Fig.~4 \cite{Fig4info}, three of the
models are now definitively ruled out.  The full class of
inflationary models can make a variety of predictions, but the
prediction of the simplest inflationary models with large
$\Lambda$, shown on the graph, fits the data beautifully.

\section{Before and After Inflation}

Research in recent years has included investigations of what
might have preceded inflation, and how an inflationary epoch
might have ended.

Soon after the first inflationary models were introduced, several
physicists \cite{Steinhardt1983,Vilenkin1983,Linde1986} realized
that once inflation began, it would in all likelihood never stop. 
Regions of space would stop inflating, forming what can be called
``pocket universes,'' one of which would contain the observed
universe.  Nonetheless, at any given moment some portion of the
universe would still be undergoing exponential expansion, in a
process called ``eternal inflation.'' In the model depicted in
Fig.~1, for example, quantum-mechanical effects compete with the
classical motion to produce eternal inflation.  Consider a region
of size $H^{-1}$, in which the average value of $\phi$ is near
(b) or (d) in the diagram.  Call the average energy density
$\rho_0$. Whereas the classical tendency of $\phi$ is to roll
slowly downward (red arrow) toward the minimum of its potential,
the field will also be subject to quantum fluctuations (green
arrows) similar to those described above. The quantum
fluctuations will give the field a certain likelihood of hopping
up the wall of potential energy rather than down it.  Over a time
period $H^{-1}$, this region will grow $e^3 \cong 20$ times its
original size.  If the probability that the field will roll up
the potential hill during this period is greater than $1/20$,
then on average the volume of space in which $\rho >
\rho_0$ increases with time \cite{Linde1990,Linde1986,Guth2000}. 
The probability of upward fluctuations tends to become large when
the initial value of $\phi$ is near the peak at (a) or high on
the hill near (d), so for most potential energy functions the
condition for eternal inflation is attainable.  In that case the
volume of the inflating region grows exponentially, and forever:
Inflation would produce an infinity of pocket universes.

An interesting question is whether or not eternal inflation makes
the big bang unnecessary: Might eternal inflation have been truly
``eternal,'' existing more or less the same way for all time, or
is it only ``eternal'' to the future once it gets started? Borde
and Vilenkin have analyzed this question (most recently, with
Guth), and have concluded that eternal inflation could not have
been past-eternal: Using kinematic arguments, they showed
\cite{BordeGuthVilenkin2003} that the inflating region must have
had a past boundary, before which some alternative description
must have applied.  One possibility would be the creation of the
universe by some kind of quantum process.

Another major area of research centers on the mechanisms by which
inflation might have ended within our observable universe.  The
means by which inflation ends have major consequences for the
subsequent history of our universe.  For one thing, the colossal
expansion during inflation causes the temperature of the universe
to plummet nearly to zero, and dilutes the density of ordinary
matter to negligible quantities.  Some mechanism must therefore
convert the energy of the scalar field, $\phi$, into a hot soup
of garden-variety matter.

In most models, inflation ends when $\phi$ oscillates around the
minimum of its potential, as in region (c) of Fig.~1. 
Quantum-mechanically, these field oscillations correspond to a
collection of $\phi$ particles approximately at rest.  Early
studies of post-inflation ``reheating'' assumed that individual
$\phi$ particles would decay during these oscillations like
radioactive nuclei.  More recently, it has been discovered
\cite{KLS,STB,Boyanovsky,Kaiser,Kofmanrev} that these
oscillations would drive resonances in $\phi$'s interactions with
other quantum fields.  Instead of individual $\phi$ particles
decaying independently, these resonances would set up collective
behavior --- $\phi$ would release its energy more like a laser
than an ordinary light bulb, pouring it extremely rapidly into a
sea of newly created particles.  Large numbers of particles would
be created very quickly within specific energy-bands,
corresponding to the frequency of $\phi$'s oscillations and its
higher harmonics.

This dramatic burst of particle creation would affect spacetime
itself, as it responded to changes in the arrangement of matter
and energy.  The rapid transfer of energy would excite
gravitational perturbations, of which the most strongly amplified
would be those with frequencies within the resonance bands of the
decaying $\phi$ field.  In some extreme cases, very
long-wavelength perturbations can be amplified during reheating,
which could in principle even leave an imprint on the CMB
\cite{metricperts}. 

\section{Brane Cosmology:  Sticking Close to Home}

Although superstring theory promises to synthesize general
relativity with the other fundamental forces of nature, it
introduces a number of surprising features --- such as the
existence of microscopic strings, rather than particles, as the
fundamental units of matter, along with the existence of several
extra spatial dimensions in the universe.  Could our observable
universe really be built from such a bizarre collection of
ingredients?

Na\"\i vely, one might expect the extra dimensions to conflict
with the observed behavior of gravity.  To be successful, string
theory, like general relativity, must reduce to Newton's law of
gravity in the appropriate limit.  In Newton's formulation,
gravity can be described by force lines that always begin and end
on masses.  If the force lines could spread in $n$ spatial
dimensions, then at a radius $r$ from the center, they would
intersect a hypersphere with surface area proportional to
$r^{n-1}$.  An equal number of force lines would cross the
hypersphere at each radius, which means that the density of force
lines would be proportional to 1/$r^{n-1}$.  For $n = 3$, this
reproduces the familiar Newtonian force law, $F \propto 1/r^2$,
which has been tested (along with its Einsteinian generalization)
to remarkable accuracy over a huge range of distances, from
astronomical scales down to less than a millimeter
\cite{Will,Hoyle}.

An early response to this difficulty was to assume that the extra
spatial dimensions are curled up into tiny closed circles rather
than extending to macroscopic distances.  Because gravity has a
natural scale, known as the Planck length, $l_P \equiv
\sqrt{\hbar G / c^3} \cong 10^{-35}$ m [where $\hbar$ is Planck's
constant divided by $2\pi$], physicists assumed that $l_P$ sets
the scale for these extra dimensions.  Just as the surface of a
soda straw would appear one-dimensional when viewed from a large
distance --- even though it is really two-dimensional --- our
space would appear three-dimensional if the extra dimensions were
``compactified'' in this way.  On scales much larger than the
radii of the extra dimensions, $r_c$, we would fail to notice
them: The strength of gravity would fall off in its usual $1/r^2$
manner for distances $r \gg r_c$, but would fall off as
$1/r^{n-1}$ for scales $r \ll r_c$ \cite{Polchinski}. The
question remained, however, what caused this compactification,
and why this special behavior affected only some but not all
dimensions.

Recently Arkani-Hamed, Dimopoulos, and Dvali \cite{largedims}
realized that there is no necessary relation between $l_P$ and
$r_c$, and that experiments only require $r_c \leq 1$ mm. 
Shortly afterward, Randall and Sundrum \cite{RS2,LRrev}
discovered that the extra dimensions could even be infinite in
extent! In the Randall-Sundrum model, our observable universe
lies on a membrane, or ``brane'' for short, of three spatial
dimensions, embedded within some larger multidimensional space. 
The key insight is that the energy carried by the brane will
sharply affect the way the gravitational field behaves.  For
certain spacetime configurations, the behavior of gravity along
the brane can appear four-dimensional (three space and one time),
even in the presence of extra dimensions.  Gravitational force
lines would tend to ``hug'' the brane, rather than spill out into
the ``bulk'' --- the spatial volume in which our brane is
embedded.  Along the brane, therefore, the dominant behavior of
the gravitational force would still be $1/r^2$.

In simple models, in which the spacetime geometry along our brane
is highly symmetric, such as the Minkowski spacetime of special
relativity, the effective gravitational field along our brane is
found to mimic the usual Einsteinian results to high accuracy
\cite{GarrigaTanaka,GKR}.  At very short distances there are
calculable (and testable) deviations from standard gravity, and
there may also be deviations for very strong gravitational
fields, such as those near black holes. There are also
modifications to the cosmological predictions of gravity. In the
usual case, when Einstein's equations are applied to a homogeneous
and isotropic spacetime, one finds $H^2 \propto \rho - k/a^2$,
where $k$ is a constant connected to the curvature of the
universe. If instead we lived on a brane embedded within one
large extra dimension, then $H^2 \propto \rho + \alpha \rho^2 -
k/a^2$, where $\alpha$ is a constant \cite{BDL}.

Under ordinary conditions, $\rho$ decreases as the universe
expands, and so the new term in the effective Einstein equations
should have minimal effects at late times in our observable
universe.  But we saw above that during an inflationary epoch,
$\rho \cong {\rm constant}$, and in these early moments the
departures from the ordinary Einsteinian case can be dramatic. 
In particular, the $\rho^2$ term would allow inflation to occur
at lower energies than are usually assumed in ordinary
(nonembedded) models, with potential energy functions that are
less flat than are ordinarily needed to sustain inflation.
Moreover, the spectrum of primordial perturbations would get
driven even closer to the scale-invariant shape, with $n_s =
1.00$ \cite{MaartensWandsBassettHeard,Langlois}.  Brane cosmology
thus leads to some interesting effects during the early universe,
making inflation even more robust than in ordinary scenarios.

\section{String Cosmology}

Although theories of extra dimensions establish a connection
between string theory and cosmology, the developments of the past
few years have pushed the connection much further.  [For reviews,
see \cite{Quevedo,LindeStringrev,Burgess}.] The union of string
theory and cosmology is barely past its honeymoon, but so far the
marriage appears to be a happy one.  Inflation, from its
inception, was a phenomenologically very successful idea that has
been in need of a fundamental theory to constrain its many
variations.  String theory, from its inception, has been a very
well-constrained mathematical theory in need of a phenomenology
to provide contact with observation.  The match seems perfect,
but time will be needed before we know for sure whether either
marriage partner can fulfill the needs of the other.  In the
meantime, ideas are stirring that have the potential to radically
alter our ideas about fundamental laws of physics. 

For many years the possibility of describing inflation in terms
of string theory seemed completely intractable, because the only
string vacua that were understood were highly supersymmetric
ones, with many massless scalar fields, called moduli, which have
potential energy functions that vanish identically to all orders
of perturbation theory.  When the effects of gravity are
included, the energy density of such supersymmetric states is
never positive. Inflation, on the other hand, requires a positive
energy density, and it requires a hill in the potential energy
function.  Inflation, therefore, could only be contemplated in
the context of nonperturbative supersymmetry-breaking effects, of
which there was very little understanding.

The situation changed dramatically with the realization that
string theory contains not only strings, but also branes, and
fluxes, which can be thought of as higher-dimensional
generalizations of magnetic fields.  The combination of these two
ingredients makes it possible to construct string theory states
that break supersymmetry, and that give nontrivial potential
energy functions to all the scalar fields. 

One very attractive idea for incorporating inflation into string
theory is to use the positions of branes to play the role of the
scalar field that drives inflation.  The earliest version of this
theory was proposed in 1998 by Dvali and Tye \cite{DvaliTye},
shortly after the possibility of large extra dimensions was
proposed in \cite{largedims}. In the Dvali-Tye model, the
observed universe is described not by a single three-dimensional
brane, but instead by a number of three-dimensional branes which
in the vacuum state would sit on top of each other.  If some of
the branes were displaced, however, in a fourth spatial
direction, then the energy would be increased.  The brane
separation would be a function of time and the three spatial
coordinates along the branes, and so from the point of view of an
observer on the brane, it would act like a scalar field that
could drive inflation.  At this stage, however, the authors
needed to invoke unknown mechanisms to break supersymmetry and to
give the moduli fields nonzero potential energy functions.

In 2003, Kachru, Kallosh, Linde, and Trivedi \cite{KKLT} showed
how to construct complicated string theory states for which all
the moduli have nontrivial potentials, for which the energy
density is positive, and for which the approximations that were
used in the calculations appeared justifiable.  These states are
only metastable, but their lifetimes can be vastly longer than
the 10 billion years that have elapsed since the big bang.  There
was nothing elegant about this construction --- the six extra
dimensions implied by string theory are curled not into circles,
but into complicated manifolds with a number of internal loops
that can be threaded by several different types of flux, and
populated by a hodgepodge of branes.  Joined by Maldacena and
McAllister, this group \cite{KKLMMT} went on to construct states
that can describe inflation, in which a parameter corresponding
to a brane position can roll down a hill in its potential energy
diagram.  Generically the potential energy function is not flat
enough for successful inflation, but the authors argued that the
number of possible constructions was so large that there may well
be a large class of states for which sufficient inflation is
achieved.  Iizuka and Trivedi \cite{IT} showed that successful
inflation can be attained by curling the extra dimensions into a
manifold that has a special kind of symmetry.

A tantalizing feature of these models is that at the end of
inflation, a network of strings would be produced
\cite{PolchinskiFD}.  These could be fundamental strings, or
branes with one spatial dimension.  The CMB data of Fig.~4 rule
out the possibility that these strings are major sources of
density fluctuations, but they are still allowed if they are
light enough so that they don't disturb the density fluctuations
from inflation.  String theorists are hoping that such strings
may be able to provide an observational window on string physics.

A key feature of the constructions of inflating states or
vacuumlike states in string theory is that they are far from
unique.  The number might be something like $10^{500}$
\cite{BPJHEP,Susskind,BPSciAm}, forming what Susskind has dubbed
the ``landscape of string theory.'' Although the rules of string
theory are unique, the low-energy laws that describe the physics
that we can in practice observe would depend strongly on which
vacuum state our universe was built upon.  Other vacuum states
could give rise to different values of ``fundamental'' constants,
or even to altogether different types of ``elementary''
particles, and even different numbers of large spatial
dimensions! Furthermore, because inflation is generically
eternal, one would expect that the resulting eternally inflating
spacetime would sample every one of these states, each an
infinite number of times.  Because all of these states are
possible, the important problem is to learn which states are
probable.  This problem involves comparison of one infinity with
another, which is in general not a well-defined question
\cite{Linde1994}.  Proposals have been made and arguments have
been given to justify them \cite{GarrigaVilenkin}, but no
conclusive solution to this problem has been found.

What, then, determined the vacuum state for our observable
universe? Although many physicists (including the authors) hope
that some principle can be found to understand how this choice
was determined, there are no persuasive ideas about what form
such a principle might take.  It is possible that inflation helps
to control the choice of state, because perhaps one state or a
subset of states expands faster than any others.  Because
inflation is generically eternal, the state that inflates the
fastest, along with the states that it decays into, might
dominate over any others by an infinite amount.  Progress in
implementing this idea, however, has so far been nil, in part
because we cannot identify the state that inflates the fastest,
and in part because we cannot calculate probabilities in any
case.  If we could calculate the decay chain of the most rapidly
inflating state, we would have no guarantee that the number of
states with significant probability would be much smaller than
the total number of possible states.

Another possibility, now widely discussed, is that {\it nothing}
determines the choice of vacuum for our universe; instead, the
observable universe is viewed as a tiny speck within a multiverse
that contains {\it every} possible type of vacuum.  If this point
of view is right, then a quantity such as the electron-to-proton
mass ratio would be on the same footing as the distance between
our planet and the sun.  Neither is fixed by the fundamental
laws, but instead both are determined by historical accidents,
restricted only by the fact that if these quantities did not lie
within a suitable range, we would not be here to make the
observations.  This idea --- that the laws of physics that we
observe are determined not by fundamental principles, but instead
by the requirement that intelligent life can exist to observe
them --- is often called the anthropic principle.  Although in
some contexts this principle might sound patently religious, the
combination of inflationary cosmology and the landscape of string
theory gives the anthropic principle a scientifically viable
framework.

A key reason why the anthropic approach is gaining attention is
the observed fact that the expansion of the universe today is
accelerating, rather than slowing down under the influence of
normal gravity.  In the context of general relativity, this
requires that the energy of the observable universe is dominated
by dark energy. The identity of the dark energy is unknown, but
the simplest possibility is that it is the energy density of the
vacuum, which is equivalent to what Einstein called the
cosmological constant.  To particle physicists it is not
surprising that the vacuum has nonzero energy density, because
the vacuum is known to be a very complicated state, in which
particle-antiparticle pairs are constantly materializing and
disappearing, and fields such as the electromagnetic field are
constantly undergoing wild fluctuations.  From the standpoint of
the particle physicist, the shocking thing is that the energy
density of the vacuum is so low.  No one really knows how to
calculate the energy density of the vacuum, but na\"\i ve estimates
lead to numbers that are about $10^{120}$ times larger than the
observational upper limit.  There are both positive and negative
contributions, but physicists have been trying for decades to
find some reason why the positive and negative contributions
should cancel, but so far to no avail.  It seems even more
hopeless to find a reason why the net energy density should be
nonzero, but 120 orders of magnitude smaller than its expected
value.  However, if one adopts the anthropic point of view, it
was argued as early as 1987 by Weinberg \cite{Weinberg} that an
explanation is at hand: If the multiverse contained regions with
all conceivable values of the cosmological constant, galaxies and
hence life could appear only in those very rare regions where the
value is small, because otherwise the huge gravitational
repulsion would blow matter apart without allowing it to collect
into galaxies. 

The landscape of string theory and the evolution of the universe
through the landscape are of course still not well understood,
and some have argued \cite{BDG} that the landscape might not even
exist.  It seems too early to draw any firm conclusions, but
clearly the question of whether the laws of physics are uniquely
determined, or whether they are environmental accidents, is an
issue too fundamental to ignore.

\section{Conclusions}

During the past decade, cosmology has unquestionably entered the
domain of high-precision science.  Just a few years ago several
basic cosmological quantities, such as the expansion parameter,
$H$, and the flatness parameter, $\Omega$, were known only to
within a factor of 2. Now new observations using WMAP, SDSS, and
the high-redshift type Ia supernovae measure these and other
crucial quantities with percent-level accuracy.  Several of
inflation's most basic quantitative predictions, including
$\Omega = 1$ and $n_s \cong 1$, may now be compared with data
that are discriminating enough to distinguish inflation from many
of its theoretical rivals.  So far, every measure has been
favorable to inflation.

Even with the evidence in favor of inflation now stronger than
ever, much work remains.  Inflationary cosmology has always been
a framework for studying the interconnections between particle
physics and gravitation --- a collection of models rather than a
unique theory.  The next generation of astronomical detectors
should be able to distinguish between competing inflationary
models, whittling down the large number of options to a preferred
few.  One important goal is the high-precision measurement of
polarization effects in the CMB, which allows the possibility of
uncovering the traces of gravity waves originating from
inflation.  Gravity waves of the right pattern would be a
striking test of inflation, and would allow us to determine the
energy density of the ``false vacuum'' state that drove
inflation. The new cosmological observations also offer
physicists one of the best resources for evaluating the latest
developments in idea-rich (but data-poor) particle theory, where
much of the current research has been aimed at the high-energy
frontier, well beyond the range of existing accelerators. Perhaps
the interface between string theory and cosmology will lead to
new predictions for the astronomers to test.  Whether such tests
are successful or not, physicists are certain to learn important
lessons about the nature of space, time, and matter.

Meanwhile, several major puzzles persist.  Now that physicists
and astronomers are confident that $\Omega = 1$ to high accuracy,
the question remains of just what type of matter and energy is
filling the universe.  Ordinary matter, such as the protons,
neutrons, and electrons that make up atoms, contributes just
$4\%$ to this cosmic balance.  Nearly one-quarter of the
universe's mass-energy is some form of ``dark matter'' ---
different in kind from the garden-variety matter we see around
us, and yet exerting a measurable gravitational tug that shapes
the way galaxies behave.  Particle physicists have offered many
candidates for this exotic dark matter, but to date no single
contender has proved fully convincing \cite{OstrikerSteinhardt}.

Even more bizarre is the dark energy now known to contribute
about $70\%$ to $\Omega$.  This dark energy is driving a
mini-inflationary epoch today, billions of years after the
initial round of inflation.  Today's accelerated expansion is far
less fast than the earlier inflationary rate had been, but the
question remains why it is happening at all.  Could the dark
energy be an example of Einstein's cosmological constant? Or
maybe it is a variation on an inflationary theme: Perhaps some
scalar field has been sliding down its potential energy hill on a
time scale of billions of years rather than fractions of a second
\cite{Kirshner}. Whatever its origin, dark energy, much like dark
matter, presents a fascinating puzzle that will keep cosmologists
busy for years to come.

\begin{acknowledgments}

The authors thank Gia Dvali, Hong Liu, Max Tegmark, Sandip
Trivedi, and Henry Tye for very helpful comments on the
manuscript.  This work was supported in part by funds provided by
the U.S. Department of Energy (D.O.E.) under cooperative research
agreement \#DF-FC02-94ER40818.

\end{acknowledgments}


\begin{thebibliography}{9999}
\bibitem{Guth1981} A. H. Guth, ``The inflationary universe:  A possible 
solution to the horizon and flatness problems,''
{\it Phys. Rev. D} {\bf 23}, 347 (1981).

\bibitem{Linde1982} A. D. Linde, ``A new inflationary universe scenario:  
A possible solution of the horizon, flatness, homogeneity, isotropy and 
primordial monopole problems,'' {\it Phys. Lett. B} {\bf 108}, 389 (1982).

\bibitem{AlbrechtSteinhardt1982} A. Albrecht, P. J. Steinhardt, 
``Cosmology for grand unified theories with radiatively induced symmetry 
breaking,'' {\it Phys. Rev. Lett.} {\bf 48}, 1220 (1982).

\bibitem{Linde1990} A. D. Linde, {\it Particle Physics and 
Inflationary Cosmology} (Harwood, Philadelphia, 1990).

\bibitem{KolbTurner1990} E. W. Kolb, M. S. Turner, {\it The Early
Universe} (Addison-Wesley, Reading, MA, 1990).

\bibitem{LiddleLyth2000} A. R. Liddle, D. H. Lyth, {\it
Cosmological Inflation and Large-Scale Structure} (Cambridge
Univ.\ Press, New York, 2000).

\bibitem{GuthBook} A. H. Guth, {\it The Inflationary 
Universe: The Quest for a New Theory of Cosmic Origins}
(Addison-Wesley, Reading, MA, 1997).

\bibitem{Tolman1932} R. C. Tolman, ``Possibilities in relativistic 
thermodynamics for irreversible processes without exhaustion of free 
energy,'' {\it Phys. Rev.} {\bf 39}, 320 (1932).

\bibitem{LythRiotto1999} D. H. Lyth, A. Riotto, ``Particle physics models 
of inflation and the cosmological density perturbation,'' {\it Phys. 
Rep.} {\bf 314}, 1 (1999) [arXiv: hep-ph/9807278].

\bibitem{Tegmark2004} M. Tegmark {\it et al.}, ``Cosmological parameters 
from SDSS and WMAP,''
{\it Phys. Rev. D} {\bf 69}, 103501 (2004) [arXiv: astro-ph/0310723].

\bibitem{Fixsen1996} D.~J.~Fixsen {\it et al.}, ``The cosmic microwave 
background spectrum from the full COBE FIRAS data set,'' {\it Astrophys.\
J.} {\bf 473}, 576 (1996) [arXiv: astro-ph/9605054].

\bibitem{WMAP2003} D. N. Spergel {\it et al.}, ``First year Wilkinson 
Microwave Anisotropy Probe (WMAP) observations:  Determination of 
cosmological parameters,'' {\it Astrophys. J. 
Suppl.} {\bf 148}, 175 (2003) [arXiv: astro-ph/0302209].

\bibitem{DensPert} For a review, see V.~F.~Mukhanov,
H.~A.~Feldman, R.~H.~Brandenberger, ``Theory of cosmological 
perturbations,'' {\it Phys.  Rep.} {\bf 215},
203 (1992) and also \cite{LiddleLyth2000}.

\bibitem{Kirshner} For a review, see R. P. Kirshner, ``Throwing light on 
dark energy,'' {\it Science} {\bf 300}, 1914 (2003).

\bibitem{Vilenkin} A. Vilenkin, E. P. S. Shellard, {\it Cosmic
Strings and other Topological Defects} (Cambridge Univ.\ Press,
New York, 1994).

\bibitem{Pen97} U.-L. Pen, U. Seljak, N. Turok, ``Power spectra in global 
defect theories of cosmic structure formation,'' {\it Phys. Rev. 
Lett.} {\bf 79}, 1611 (1997) [arXiv: astro-ph/9704165].

\bibitem{PolchinskiFD} For a review, see J.~Polchinski, ``Introduction to 
cosmic F- and D-strings,'' arXiv: hep-th/0412244.

\bibitem{Fig4info} We thank Max Tegmark for providing this graph,
which shows the most precise data points for each range of $\ell$
from recent observations, as summarized in \cite{Tegmark2004}.
The cosmic string prediction is taken from \cite{Pen97}. The
other curves were all calculated for $n_s=1$, $\Omega_{\rm
baryon}=0.05$, and $H$ = 70 km s$^{-1}$ Mpc$^{-1}$, with the
remaining parameters fixed as follows.  ``Inflation with
$\Lambda$'': $\Omega_{\rm DM} \hbox{(dark matter)} =0.23$,
$\Omega_\Lambda = 0.72$, and optical depth parameter $\tau =
0.17$; ``Inflation without $\Lambda$'': $\Omega_{\rm DM}=0.95$,
$\Omega_\Lambda = 0$, $\tau = 0.06$; ``Open universe'':
$\Omega_{\rm DM}=0.25$, $\Omega_\Lambda = 0$, $\tau = 0.06$.  The
1-SD error bars include both observational uncertainty and
``cosmic variance,'' the intrinsic quantum uncertainty in the
predictions, as calculated from the ``Inflation with $\Lambda$''
model. With our current ignorance of the underlying physics, none
of these theories predicts the overall amplitude of the
fluctuations; the ``Inflation with $\Lambda$'' curve was
normalized for a best fit, and the others were normalized
arbitrarily.

\bibitem{Steinhardt1983} P. J. Steinhardt, ``Natural inflation,''
in {\it The Very Early Universe}, G. W. Gibbons, S. W. Hawking, S. T. C.  
Siklos, Eds. (Cambridge Univ.\ Press, New York, 1983), p. 251.

\bibitem{Vilenkin1983} A. Vilenkin, ``The birth of inflationary 
universes,'' {\it Phys. Rev. D} {\bf 27}, 2848 (1983).

\bibitem{Linde1986} A. D. Linde, ``Eternally existing self-reproducing 
chaotic inflationary universe,'' {\it Phys. Lett. B} {\bf 175},
395 (1986).

\bibitem{Guth2000} For a review, see A. H. Guth, ``Inflation and eternal 
inflation,'' {\it Phys. Rep.} {\bf 333-334}, 555 (2000) [arXiv: 
astro-ph/0002156].

\bibitem{BordeGuthVilenkin2003} A. Borde, A. H. Guth, 
A. Vilenkin, ``Inflationary space-times are incomplete in past 
directions,'' {\it Phys. Rev. Lett.} {\bf 90}, 151301 (2003) [arXiv: 
gr-qc/0110012].

\bibitem{KLS} L. A. Kofman, A. D. Linde, A. A. Starobinsky, ``Reheating 
after inflation,'' {\it Phys. Rev. Lett.} {\bf 73}, 3195 (1994) [arXiv: 
hep-th/9405187].

\bibitem{STB} Y. Shtanov, J. H. Traschen, R. H. Brandenberger,
``Universe reheating after inflation,'' {\it Phys. Rev. D} {\bf 51}, 5438 
(1995) [arXiv: hep-ph/9407247].

\bibitem{Boyanovsky} D. Boyanovsky, H. J. de Vega, R. Holman, D. S. Lee, 
A. Singh, ``Dissipation via particle production in scalar field 
theories,'' , {\it Phys. Rev. D} {\bf 51}, 4419 (1995) [arXiv: 
hep-ph/9408214].

\bibitem{Kaiser} D. I. Kaiser, ``Post-inflation reheating in an expanding 
universe,'' {\it Phys. Rev. D} {\bf 53}, 1776 (1996) [arXiv: 
astro-ph/9507108].

\bibitem{Kofmanrev} For a review, see L. A. Kofman, ``Preheating after 
inflation,'' in {\it COSMO-97: Proceedings}, L.\ Roszkowski, Ed. (World 
Scientific, Singapore, 1998), pp.~312--321 [arXiv: astro-ph/9802221].

\bibitem{metricperts} B. A. Bassett, D. I. Kaiser, 
R. Maartens, ``General relativistic effects in preheating,'' {\it 
Phys. Lett. B} {\bf 455}, 84 (1999) [arXiv: hep-ph/9808404].

\bibitem{Will} Clifford Will, {\it Theory and Experiment in
Gravitational Physics} (Cambridge Univ.\ Press, New York, ed.~2,
1993).

\bibitem{Hoyle} C. D.  Hoyle {\em et al.}, ``Sub-millimeter tests of the 
gravitational inverse-square law,'' {\it Phys. Rev. D}
{\bf 70}, 042004 (2004) [arXiv: hep-ph/0405262].

\bibitem{Polchinski} See, for example, J. Polchinski, {\em String
Theory} (Cambridge Univ.\ Press, New York, 1998), vol. 1.

\bibitem{largedims} N. Arkani-Hamed, S. Dimopoulos, 
G. Dvali, ``The hierarchy problem and new dimensions at a millimeter,''
{\it Phys.  Lett. B} {\bf 429}, 263 (1998) [arXiv: hep-ph/9803315].

\bibitem{RS2} L. Randall, R. Sundrum, ``An alternative to 
compactification,'' {\it Phys. Rev. Lett.} {\bf
83}, 4690 (1999) [arXiv: hep-th/9906064]. 

\bibitem{LRrev} For a review, see L. Randall, ``Extra dimensions and 
warped geometries,'' {\it Science} {\bf 296}, 1422 (2002).

\bibitem{GarrigaTanaka} J. Garriga, T. Tanaka, ``Gravity in the
Randall-Sundrum brane world,'' {\it Phys. Rev. Lett.} {\bf 84},
2778 (2000) [arXiv:  hep-th/9911055].

\bibitem{GKR}  S. B. Giddings, E. Katz, L. Randall, ``Linearized gravity 
in brane backgrounds,'' {\it J. High Energy Phys.} {\bf 3}, 23 (2000) 
[arXiv: hep-th/0002091].

\bibitem{BDL} P. Binetruy, C. Deffayet, D. Langlois, ``Non-conventional 
cosmology from a brane universe,'' {\it Nucl. Phys. B} {\bf 565}, 269 
(2000) [arXiv: hep-th/9905012].

\bibitem{MaartensWandsBassettHeard} R. Maartens, D. Wands, B. A.
Bassett, I. P. C. Heard, ``Chaotic inflation on the brane,''
{\it Phys. Rev. D} {\bf 62}, 041301 (2000) [arXiv: hep-ph/9912464].

\bibitem{Langlois} For a review, see D. Langlois, ``Gravitation and 
cosmology in brane-worlds,'' arXiv: gr-qc/0410129.

\bibitem{Quevedo} F. Quevedo, ``Lectures on string/brane cosmology,''
{\it Class. Quant. Grav.} {\bf 19}, 5721 (2002) [arXiv: hep-th/0210292].

\bibitem{LindeStringrev} A. D. Linde, ``Prospects of inflation,'' arXiv: 
hep-th/0402051.

\bibitem{Burgess} C. P. Burgess, ``Inflationary string theory?,'' {\it 
Pramana} {\bf 63}, 1269 (2004) [arXiv: hep-th/0408037].

\bibitem{DvaliTye} G.~R.~Dvali, S.-H. H. Tye, ``Brane inflation,'' {\it 
Phys. Lett. B} {\bf 450}, 72 (1999) [arXiv: hep-ph/9812483].

\bibitem{KKLT} S.~Kachru, R.~Kallosh, A.~Linde, S.~P.~Trivedi, ``De Sitter 
vacua in string theory,'' {\it Phys. Rev. D} {\bf 68}, 046005 (2003) 
[arXiv: hep-th/0301240].

\bibitem{KKLMMT} S.~Kachru, R. Kallosh, A. Linde, J. Maldacena, L. 
McAllister, and S. P. Trivedi, ``Towards inflation in string theory,'', 
{\it J. Cosmol.\ Astropart.\ Phys.\ }{\bf 0310}, 013 (2003) [arXiv: 
hep-th/0308055].

\bibitem{IT} N. Iizuka, S.~P.~Trivedi, ``An inflationary model in string 
theory,'' {\it Phys. Rev. D} {\bf 70}, 043519 (2004) [arXiv: 
hep-th/0403203].

\bibitem{BPJHEP} R.~Bousso, J.~Polchinski, ``Quantization of four-form 
fluxes and dynamical neutralization of the cosmological constant,''
{\it J. High Energy Phys.\ }{\bf 0006}, 006 (2000) [arXiv: 
hep-th/0004134].

\bibitem{Susskind}  L.~Susskind, ``The anthropic landscape of string 
theory,'' arXiv: hep-th/0302219.

\bibitem{BPSciAm} R.~Bousso, J.~Polchinski, ``The string theory 
landscape,'' {\it Sci. Am.} {\bf 291}, 60 (Sept.\ 2004).

\bibitem{Linde1994} A. D. Linde, D. Linde, A. Mezhlumian, ``Do we
live in the center of the world?,'' {\it Phys. Lett. B} {\bf
345}, 203--210 (1995) [arXiv: hep-th/9411111].

\bibitem{GarrigaVilenkin} See, for example, J.~Garriga,
A.~Vilenkin, ``A prescription for probabilities in eternal inflation,''
{\it Phys.\ Rev.\ D} {\bf 64}, 023507 (2001) [arXiv: gr-qc/0102090], and
also \cite{TegmarkLandscape}.

\bibitem{TegmarkLandscape} M.~Tegmark, ``What does inflation really 
predict?,'' arXiv: astro-ph/0410281.

\bibitem{Weinberg} S.~Weinberg, ``Anthropic bound on the cosmological 
constant,'' {\it Phys. Rev. Lett.} {\bf 59}, 2607 
(1987).

\bibitem{BDG} T.~Banks, M.~Dine, E.~Gorbatov, ``Is there a string theory 
landscape?,'' {\it J. High Energy Phys.\ }{\bf 0408}, 058 (2004) [arXiv: 
hep-th/0309170].

\bibitem{OstrikerSteinhardt} For a review, see J. P. Ostriker, P.
Steinhardt, ``New light on dark matter,'' {\it Science} {\bf 300}, 1909 
(2003) [arXiv: astro-ph/0306402].

\end{thebibliography}
\end{document}